\documentclass{svmult}

\usepackage{makeidx}     
\usepackage{graphicx}    
\usepackage{multicol}    

\makeindex             

\begin{document}

\title*{Time-scale dependence of correlations among foreign currencies
\vspace*{1.5cm}
}
\small\author{Takayuki Mizuno\inst{1},Shoko Kurihara\inst{1},
Misako Takayasu\inst{2}, and Hideki Takayasu\inst{3}}
\institute{Department of Physics, Faculty of Science and Engineering, Chuo 
University, 1-13-27  Kasuga, Bunkyo-ku, Tokyo 112-8551, Japan
\and Department of Complex Systems, Future University-Hakodate, 116-2 
Kameda-Nakano, Hakodate 041-0803, Japan
\and Sony Computer Science Laboratories, 3-14-13 Higashi-Gotanda, 
Shinagawa-ku, Tokyo, Japan}
%
%
\maketitle
\bigskip
\textbf{Summary. }For the purpose of elucidating the correlation among 
currencies, we analyze daily and high-resolution data of foreign exchange 
rates. There is strong correlation for pairs of currencies of geographically 
near countries. We show that there is a time delay of order less than a 
minute between two currency markets having a strong cross-correlation. The 
cross-correlation between exchange rates is lower in shorter time scale in 
any case. As a corollary we notice a kind of contradiction that the direct 
Yen-Dollar rate significantly differs from the indirect Yen-Dollar rate 
through Euro in short time scales. This result shows the existence of 
arbitrage opportunity among currency exchange markets.

\vspace{1cm}

\noindent
\textbf{Key words.} Foreign exchange, cross-correlation, interaction.

\vspace{0.3cm}

\section{Introduction}
\label{sec:1}
It is widely known that foreign exchange rates sometimes show very violent 
fluctuations characterized by power law distributions [T. Mizuno][K. Matia]. 
A very important report appeared recently that arbitrage chances exist by 
considering three currency markets simultaneously [Y. Aiba]. This result is 
due to the fact that the distribution of return of rotation transaction of 
three currencies show a similar fat-tail. As statistical laws among foreign 
exchange rates in the short time scale are not known sufficiently, we 
investigate correlations among foreign exchange rates in this paper.

\noindent
\textbf{Table.1} The set of Daily data for 25 kinds of foreign exchange we 
examined.

\smallskip
\noindent
\begin{tabular}{llll}
\hline
\textbf{Base Currency :} & U.S. Dollar \\
\textbf{Target Currencies } & \textbf{:} \\
\ Australian Dollar, & Brazilian Real, & British Pound, & Canadian Dollar, \\
\ Chilean Peso, & Colombian Peso, & Czech Koruna, & Euro, \\
\ Hungarian Forint, & Indonesian Rupiah, & Japanese Yen, & Mexican Peso, \\ 
\ New Zesland Dollar, & Norwegian Kroner, & Peruvian New Sole, 
& Philippines Peso, \\
\ Polish Zloty, & Russian Ruble, & Singapre Dollar, & Slovakian Koruna, \\
\ South African Rand, & South Korean Won, & Swedish Krona, & Swiss Franc, \\
\ Thai Baht \\
\hline
\end{tabular}

\vspace{1cm}
Correlations among financial stocks in markets have been actively discussed. 
The strength of synchronization among stocks was analyzed by using the 
ultrametric spaces and they discussed about the portfolio of stock [R. N. 
Mantegna, H. E. Stanley]. The interaction of companies was investigated by 
analyzing return of many stocks and a directed network of influence among 
companies was defined [L. Kullmann]. It will be noticed in the present paper 
that the correlation among foreign exchange rates resembles the stock case.

 In the following section we discuss the cross-correlation between exchange 
rates with no time difference to show the relation between synchronization 
of exchange rate. Then, we show the maximum correlation between exchange 
rates is observed with nonzero time shift, namely, the direction of 
influence is discussed.

\section{The correlation among foreign exchange rates}
\label{sec:2}
We first analyze a set of daily data provided by Exchange Rate Service [ERS] 
for 25 exchange rates for about 3 years from January `99 to August `01 as 
listed in Table 1. We first estimate cross-correlation functions for these 
exchange rates measured by USD (United States Dollar). The largest 
correlation value (=0.95) is observed for a pair of CHF(Swiss Franc)/USD and 
EUR(Euro)/USD. As demonstrated in Fig.1 it is evident that these exchange 
rates are remarkably synchronized. There are cases with negative correlation 
values as found for the pair of MXP(Mexican-Peso)/USD and CHF(Swiss 
Franc)/USD whose correlation value is |0.23. It is known that correlations 
between geographically closer currencies tend to have larger correlation and 
there exist key currencies in each area, for example, Euro for West Europe, 
Yen for Asia, Hungarian Forint for East Europe and Australian Dollar for 
Oceania [H. Takayasu and M. Takayasu].

Although we can find such large correlations among currencies in daily data, 
we can expect low correlations in short time scale as it is common that 
dealers in major banks tend to work with a single foreign exchange market. 
In order to clarify this tendency we examine 
tick-by-tick data provided by Reuters for 
about 4 months from March `02 to July `02. In Fig.2 we plot the correlation 
value as a function of coarse-grained time-scale for a pair of CHF/USD and 
EUR/USD. From this figure 

\newpage

\begin{figure}[htbp]
\centerline{\includegraphics[width=9.4cm,height=5.4cm]{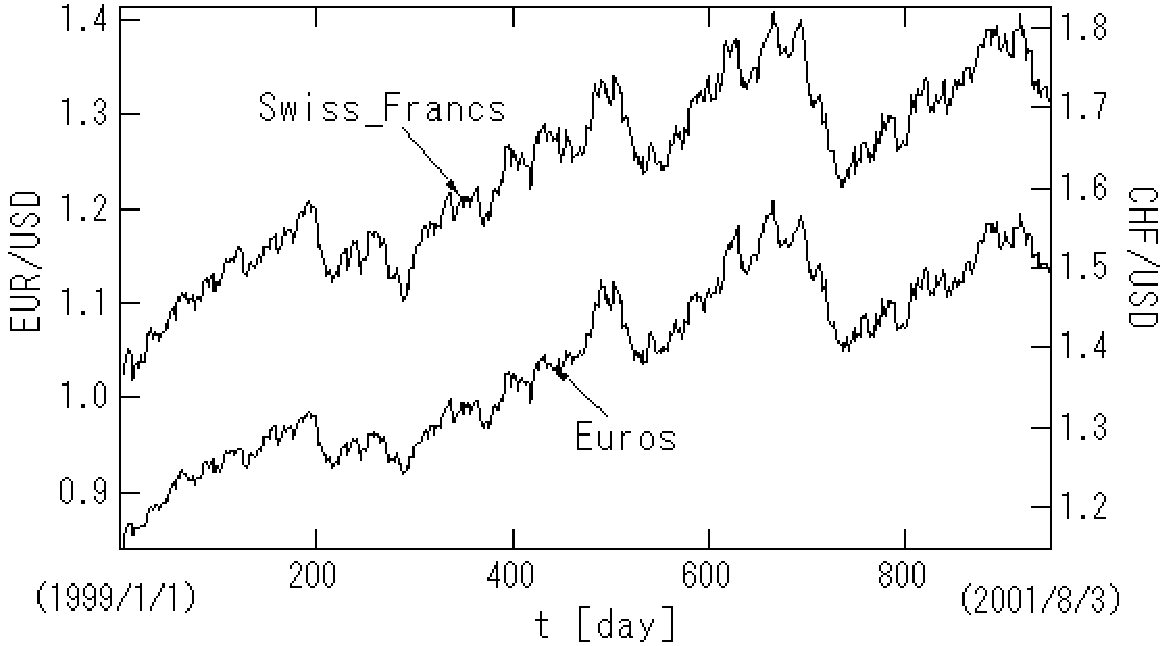}}
\center{
\textbf{Fig.1} Daily time series of CHF(Swiss-Franc)/Dollar and Euro/Dollar.
}
\end{figure}

\begin{figure}[htbp]
\centerline{\includegraphics[width=8.4cm,height=5.6cm]{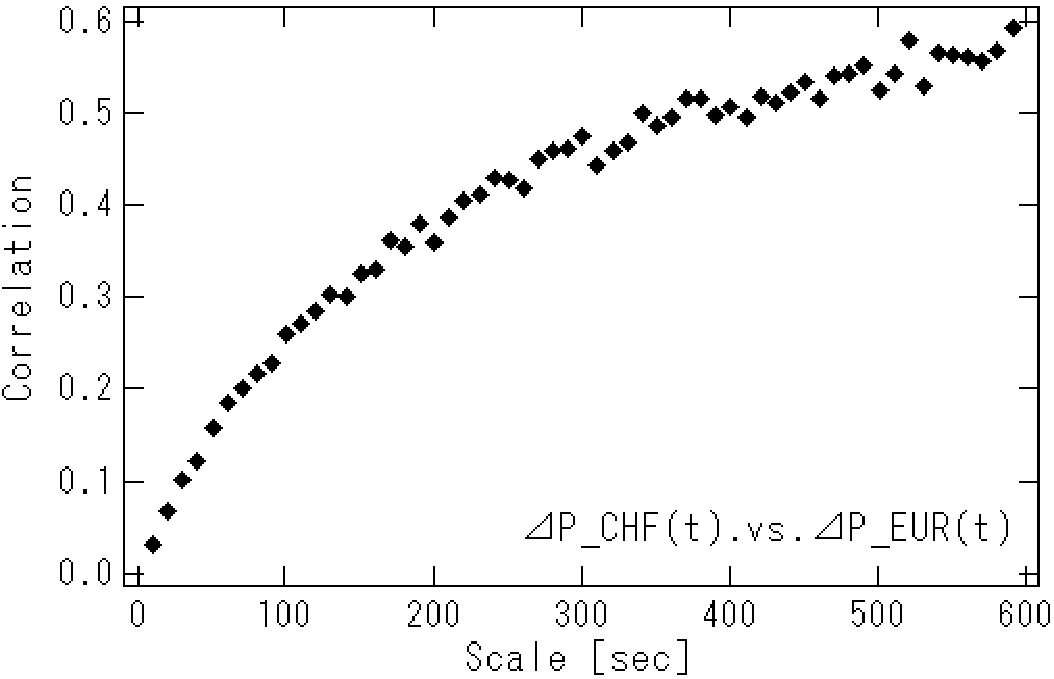}}
\center{
\textbf{Fig.2} The correlation value as a function of observation 
time-scale.
}
\end{figure}

\noindent
we notice that the correlation vanishes if we 
observe the high-resolution data with the precision of seconds and the 
correlation value is about 0.5 in the time scale of 5 minutes (300 sec.). 
From these results it is understood that these two currency markets, CHF/USD 
and EUR/USD, are working independently in very short time-scale.

\newpage

\begin{figure}[htbp]
\centerline{\includegraphics[width=8.4cm,height=5.4cm]{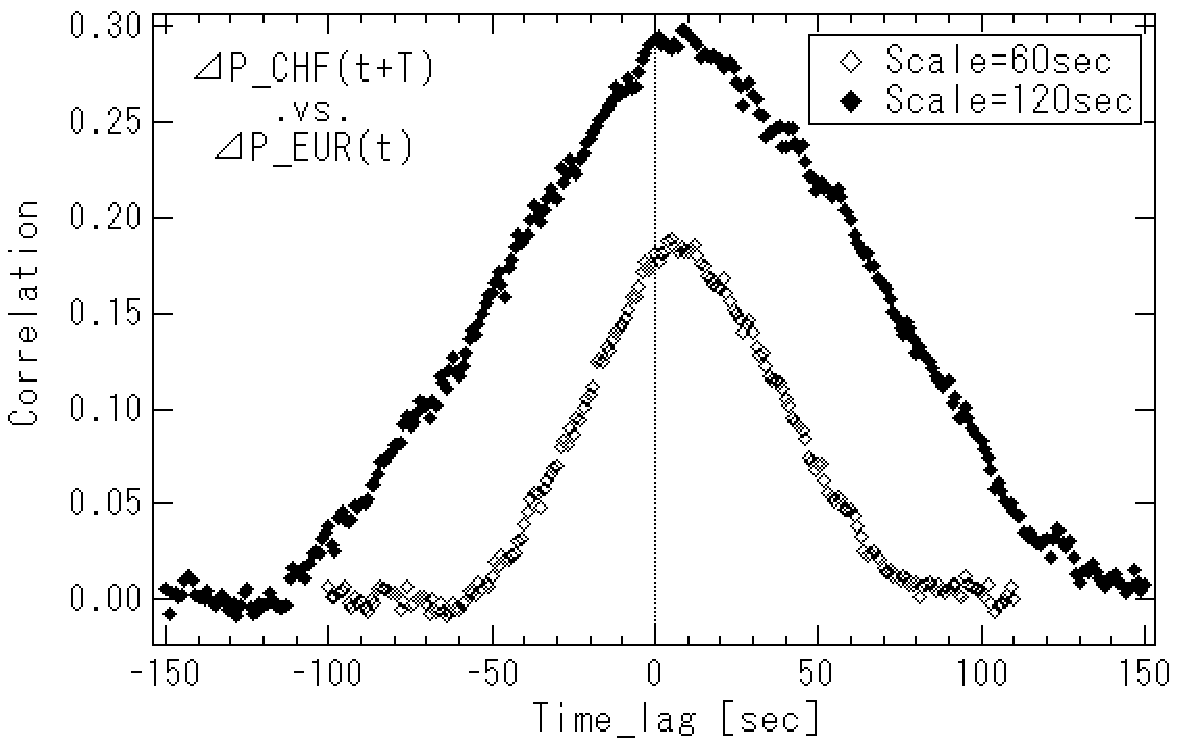}}
\center{\textbf{Fig.3} The cross-correlation with a time shift of CHF/USD and 
EUR/USD.
}
\end{figure}

\section{The cross-correlation with a time shift}
\label{sec:3}
In order to clarify the nature of short time interaction among currencies we 
calculate the cross-correlation with a time shift, namely, we observe the 
correlation of these two markets with a time difference by the following 
equation, 

\vspace{0.2cm}
\begin{equation}
\label{eq1}
C(dt) = \frac{\left\langle {dp_A (t) \cdot dp_B (t + dt)} \right\rangle - 
\left\langle {dp_A (t)} \right\rangle \left\langle {dp_B (t + dt)} 
\right\rangle }{\sigma _A \cdot \sigma _B },
\end{equation}
\vspace{0.2cm}

\noindent
where, $dp_A (t)$ is the rate change in the market $A$ at time $t$, $dp_B (t 
+ dt)$ is the rate change in the market $B$ at time $t + dt$, $\sigma $ is 
the standard deviation of rate changes in each market. In Fig.3, we show the 
correlation value between CHF/USD at time $t + dt$ and EUR/USD at time $t$ 
as a function of time difference $dt$. Here, we show two plots for different 
coarse-graining time-scales, 60 sec. and 120 sec.. In both cases it is found 
that the largest value of correlation is observed around $dt = 10$ seconds, 
which implies that in an average sense the EUR/USD market is going about 10 
second ahead and the CHF/USD market is following it.

\section{Currency correlation in short time scale}
\label{sec:4}
Here, we discuss about value of currency correlation in each foreign 
exchange market in a short time scale. From Fig.2, it is noticed that the 
correlation between foreign exchanges is very low in a short time scale. 
Namely, each exchange rate is changing rather independently. In order to 
clarify this property, we analyze the 
exchange rate of Yen-Dollar by analyzing a set of tick-by-tick data provided 
by CQG for about 2 years from February `99 to March `02.

\begin{figure}[htbp]
\centerline{\includegraphics[width=8.2cm,height=4.8cm]{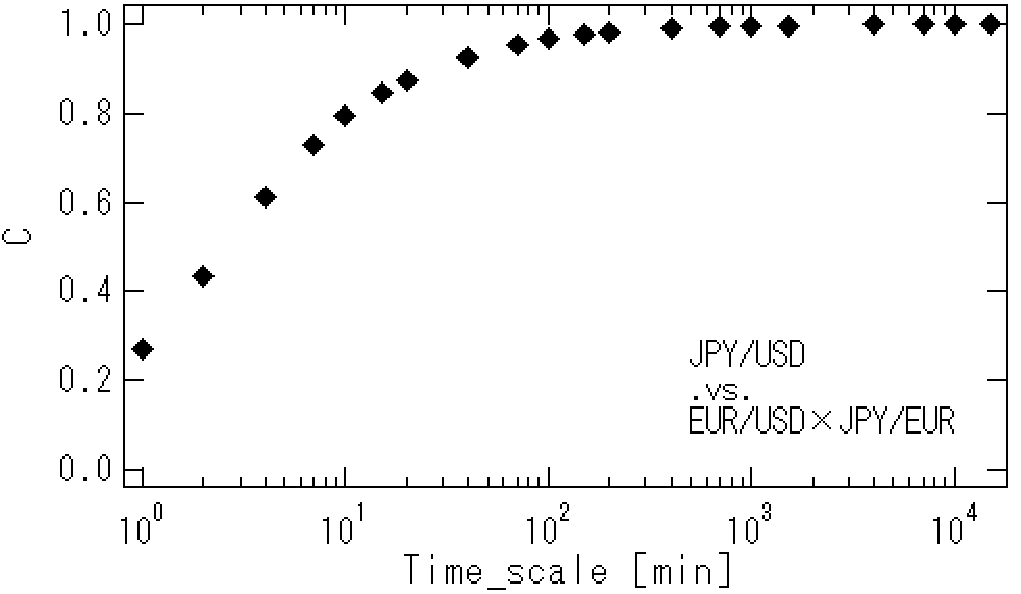}}
\textbf{Fig.4} The cross-correlation of JPY(Japanese Yen)/USD(U. S. Dollar) 
and {\{}EUR(Euro)/USD{\}}$\times ${\{}JPY/EUR{\}}.
\end{figure}

\begin{figure}[htbp]
\centerline{\includegraphics[width=8.6cm,height=4.8cm]{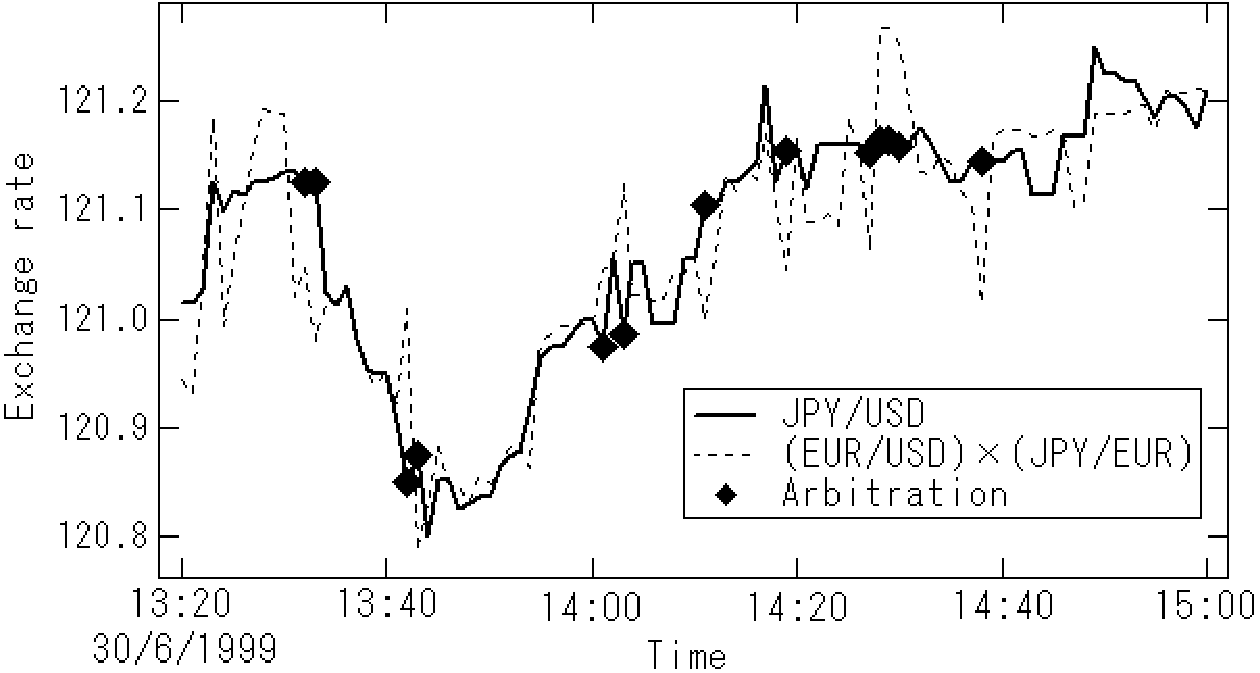}}
\textbf{Fig.5} Triangular arbitrage opportunity. A full line is JPY/USD and 
a dashed line is {\{}EUR(Euro)/USD{\}}$\times ${\{}JPY/EUR{\}}. Triangular 
arbitrage occurred at the time of Ÿ.
\end{figure}

We introduce two definitions of Yen-Dollar rate: One definition of JPY/USD 
is the usual transaction rate and the other JPY/USD is defined through Euro 
as {\{}EUR/USD{\}}$\times ${\{}JPY/EUR{\}}. Here, all exchange rates are 
given by the middle rate (=(Bid rate + Ask rate)/2). In Fig.4 we plot the 
cross-correlation value of these exchange rates in different time scales. 
The correlation value is not unity in the time scale less than about 1 hour. 
This result clearly shows that the value of a currency differs in different 
markets in the time scale less than an hour. This is a kind of 
self-contradiction of markets causing the occurrence of triangular arbitrage 
opportunity as shown in Fig.5 [Y. Aiba].

\section{Discussion}
\label{sec:5}
We have clarified the detail properties of correlation among foreign 
exchanges; the short time correlation is generally very small even between 
the pair of currencies showing large correlation in daily data. Within the 
time scale of a minute we can observe the direction of influence from one 
currency market to others. In very short time scale we can find 
contradiction of exchange rates between JPY/USD and {\{}EUR/USD{\}}$\times 
${\{}JPY/EUR{\}}. For example, if you observe carefully the two rates, 
JPY/USD and {\{}EUR/USD{\}}$\times ${\{}JPY/EUR{\}}, then you can buy Yen 
cheaper in one market and can sell it with higher rate in the other market 
even taking into account the effect of the spread (=Ask rate -- Bid rate) 
that is about 0.05{\%}. Although, no time lag is considered regarding the 
actual transactions, it is now clear how triangular arbitrage opportunity 
appears.

\bigskip
\noindent
\textbf{Acknowledgement}

\medskip
\noindent
We would like to thank Mr. Hiroyuki Moriya of Oxford Financial Education for 
providing us with the CQG data, Prof. Tohru Nakano and Prof. Mitsuo Kono for 
stimulus discussions.

\bigskip
\noindent
\textbf{References}

\medskip
\noindent
Mizuno, T., Kurihara, S., Takayasu, M., Takayasu, H., Analysis of 
high-resolution foreign exchange data of USD-JPY for 13 years, Physica A, in 
press.

\noindent
Matia, K., et al (2002) Non-Levy Distribution of Commodity Price 
Fluctuations, Phys. Rev. E \textbf{66}, 045103.

\noindent
Aiba, Y., et al (2002) Triangular arbitrage as an interaction among foreign 
exchange rates, Physica A \textbf{310}, 467-479.

\noindent
Mantegna, R. N., Stanley, H. E., (2000) An Introduction to Econophysics: 
Correlation and Complexity in Finance, Cambridge University Press, 
Cambridge.

\noindent
Kullmann, L., Kertesz, J., Kaski, K., (2002) Time-dependent 
cross-correlations between different stock returns: A directed network of 
influence, Phys. Rev. E \textbf{66}, 026125.

\noindent
Exchange Rate Service: http://pacific.commerce.ubc.ca/xr/ .

\noindent
Takayasu, H., Takayasu, M., (2001) Econophysics-Toward Scientific 
Reconstruction of Economy, Nikkei, Tokyo. (in Japanese)

\printindex
\end{document}